\documentclass[]{bytedance_seed}

\usepackage[toc,page,header]{appendix}
\usepackage{amsmath,amsthm,amsfonts,amssymb}
\usepackage{braket}
\usepackage{mathtools}
\usepackage{bbm}
\usepackage{dsfont}
\usepackage{physics}
\usepackage{algorithm}
\usepackage{algorithmicx}
\usepackage{algpseudocode}
\usepackage{enumerate}
\usepackage{float}
\usepackage{flafter}
\usepackage{xspace}
\usepackage{array}

\definecolor{darkgreen}{RGB}{0,200,0}

\newif\ifshowcomments
\showcommentsfalse
\ifshowcomments
  \newcommand{\DD}[1]{{\color{blue}{[DD: #1]}}}
  \newcommand{\ZC}[1]{{\color{purple}{[ZS: #1]}}}
  \newcommand{\HP}[1]{{\color{red}{[HP: #1]}}}
  \newcommand{\rj}[1]{{\color{brown}{[RJ: #1]}}}
  \newcommand{\TL}[1]{{\color{darkgreen}{[TL: #1]}}}
\else
  \newcommand{\DD}[1]{}
  \newcommand{\ZC}[1]{}
  \newcommand{\HP}[1]{}
  \newcommand{\rj}[1]{}
  \newcommand{\TL}[1]{}
\fi

\newcolumntype{L}[1]{>{\raggedright\arraybackslash}p{#1}}

\title{A Unified Generative Framework for Scalable Chemical Reaction Network Exploration}

\author[1,*]{Zechang Sun}
\author[1,*]{Chenxi Hu}
\author[1,*]{Kailai Lin}
\author[1]{Jin Li}
\author[1]{Changsu Cao}
\author[1,\dagger]{Dingshun Lv}
\author[2,\dagger]{Ji Chen}
\author[1]{Weiluo Ren}
\author[1,\dagger]{Hung Q. Pham}

\affiliation[1]{ByteDance Seed}
\affiliation[2]{Peking University}

\contribution[*]{These authors contributed equally to this work}
\contribution[\dagger]{Corresponding authors}

\abstract{
Chemical reaction networks (CRNs) are crucial for understanding reaction mechanisms and guiding chemical synthesis, yet the computational exploration remains limited by the combinatorial growth of chemical space, the reliability of reaction path screening, and the cost of evaluating thermodynamic and kinetic properties. 
Here, we present ByteCRN, an end-to-end framework for computational CRN exploration that combines chemically informed reaction enumeration with generative transition state modeling.  
A key component of our framework is a generative rectified flow architecture for both transition state generation and reaction validation, where it maps reactant-product pairs to candidate transition state structures and verifies connectivity by mapping back to reactants and products.
This unified generative strategy replaces the most expensive steps of conventional computational workflows, namely iterative transition state search and intrinsic reaction coordinate validation, within a complete CRN construction pipeline. 
ByteCRN delivers a 10--100-fold acceleration over traditional workflows while maintaining high predictive fidelity for individual reactions. 
At the network scale, it effectively prunes $\sim$70-90\% of the enumerated reactions, streamlining the exploration of complex reaction space.
Its utility is illustrated through the discovery of novel pathways involving cyanoacetaldehyde and the successful modeling of the challenging $\gamma$-ketohydroperoxide network, demonstrating a practical, scalable approach to autonomous chemical exploration.
}

\date{\today}
\correspondence{Dingshun Lv at \email{lvdingshun@bytedance.com}, Ji Chen at \email{ji.chen@pku.edu.cn}, Hung Q. Pham at \email{hung.pham@bytedance.com}}
\begin{document}
\maketitle

\section{Introduction}

% Paragraph 1: What are CRNs and why are they important for chemistry
Chemical reaction networks (CRNs) describe how chemical systems evolve through interconnected elementary reactions, providing a network-level blueprint that links molecular structure, thermodynamics, and kinetics.~\cite{unsleber_exploration_2020, WEN2023} Beyond individual reactions, CRNs reveal how competing pathways, rate-limiting steps, branching behavior, and time-dependent dynamics collectively shape observable outcomes such as selectivity, yield, and stability.~\cite{proppe_mechanism_2019, motagamwala_microkinetic_2021, bensberg_uncertainty-aware_2024, MORANDI2026} Therefore, reliable CRNs underpin a wide range of applications, including catalyst design, materials degradation analysis, combustion chemistry, and autonomous reaction optimization in closed-loop laboratories.~\cite{ULISSI2017,liu_reaction_2021,BLAU2021,steiner_autonomous_2022,MCDERMOTT2021,szymanski_autonomous_2023,leonov_integrated_2024} By identifying the reactions and species that govern the network behavior, CRNs enable mechanistic interpretation and provide a foundation for predictive chemistry.

% Paragraph 2: Historically, CRNs are probed experimentally. But experiments see too little. 
Despite their central role in understanding and controlling chemical reactivity, CRNs are only partially accessible experimentally. Experiments typically resolve a small number of kinetically dominant pathways and detectable intermediates, which is only a sparse projection of the underlying network.~\cite{ji_autonomous_2021, Steiner2024} Short-lived intermediates, low-probability channels, and parallel reaction routes are often missed. Yet, these hidden species and channels can become decisive under changing experimental conditions that reshape the free-energy landscape and activate otherwise dormant routes.~\cite{park_trends_2020, motagamwala_microkinetic_2021, zhao_deep_2022, bensberg_uncertainty-aware_2024, DING2026}

% Paragraph 3: Computation can recover hidden species/rxns, but traditional workflows generate too much and validate too slowly
First-principles computation offers a route to recover this hidden network systematically from molecular physics.~\cite{https://doi.org/10.1002/anie.202011941, baiardi_expansive_2022, unsleber_chemoton_2022, zhang_automated_2023, Steiner2024, bensberg_uncertainty-aware_2024} But rather than observing too few pathways, computational exploration rapidly generates too many.~\cite{kim_efficient_2018, unsleber_exploration_2020, WEN2023, zhao_comprehensive_2023, bensberg_uncertainty-aware_2024} Even moderately sized molecular systems contain a combinatorial number of possible intermediates and reaction channels, making exhaustive first-principles exploration impractical.~\cite{VERNUCCIO2019} The central challenge is therefore not only to enumerate possible transformations, but to identify the small subset that is physically valid and kinetically relevant. 
Existing workflows typically separate CRN construction into two stages: expansive candidate generation followed by quantum chemistry validation. Rule-, template-, and graph-based methods can efficiently enumerate possible reactions using prescribed bond-breaking and bond-forming rules, enabling broad exploration of reaction space.~\cite{RMG2016,SIMM2017,RDCHIRAL2019,YARP2021,SYNKIT2025,FLOWER2025}
% A range of computational strategies has been developed to construct chemical reaction networks, from rule-based reaction enumeration \cite{RMG2016, RDCHIRAL2019, SYNKIT2025} to explicit exploration of potential energy surfaces \cite{SHANGCHENG2013}. Template- and graph-based approaches efficiently enumerate possible reactions using predefined bond-breaking and bond-forming rules~\cite{Zhao2021YARP,YARP2023,doi:10.1021/acs.jctc.7b00945}, but 
However, enumeration alone does not establish whether a proposed reaction is thermodynamically accessible, kinetically relevant, or connected via a valid reaction pathway. As a consequence, assessing the physical relevance of enumerated reactions ultimately hinges on computationally intensive first-principles validation, in which transition state (TS) search and intrinsic reaction coordinate (IRC) calculations are required to obtain activation energies.~\cite{fukui_path_1981, gonzalez_reaction_1990, unsleber_chemoton_2022, zhao_comprehensive_2023, choi_prediction_2023} Together, these TS-centered validation steps form the central bottleneck limiting automated CRN exploration.

% Paragraph 4: GenAI accelerates TS prediction on the level of individual rxns, but rarely solve network-level CRN construction. It has other issues that limit performance as well. 
This challenge has motivated increasing interest in data-driven approaches, including machine-learned potentials~\cite{NEURALNEB2022, TRANSITION1X2022, NEURALIRC2024, ramakrishnan_implicit_2025, ZHAOQIYUAN2025} and generative artificial intelligence models~\cite{OAREACTDIFF2023, choi_prediction_2023, TSDIFF2024, REACTOT2024, ZHAOQIYUAN2025, galustian_goflow_2025, hayashi_generative_2025, FLOWER2025}, to accelerate transition state identification and reaction energetics evaluation. By learning from existing reaction data, such methods can propose transition state structures or estimate activation barriers at significantly reduced computational cost.~\cite{NEURALNEB2022,NEURALIRC2024,REACTOT2024} 
Among these emerging methodologies, generative models are particularly attractive because they can learn high-dimensional data distributions and generate candidate reaction structures at low computational cost. Although previous studies have demonstrated their utility in transition state searches for individual reactions~\cite{TSGAN2021,OAREACTDIFF2023,REACTOT2024,TSDIFF2024, RitS2026}, they are seldom integrated into the broader context of automated reaction network construction.~\cite{WEN2023, OAREACTDIFF2023, ZHAOQIYUAN2025, FLOWER2025} Furthermore, the out-of-distribution generalization of generative models in CRN exploration remains hugely underexplored, despite its critical importance for practical applications.

%These approaches have enabled substantial progress in reaction-level prediction, but most still operate as isolated accelerators within conventional workflows. 

%Only a limited number of generative models have been developed with the explicit goal of supporting reaction network construction. As a result, generative models often provide improved TS guesses rather than a coupled framework for reaction enumeration, connectivity validation, and CRN assembly. 
% Most existing approaches are trained on relatively small datasets composed primarily of isolated elementary reactions, often curated for specific reaction classes or molecular motifs~\cite{Schreiner2022}. As a result, their applicability beyond these narrow settings remains uncertain, particularly when extending to reaction networks involving different chemical environments, larger structural diversity, or multi-step reactions. 
%Furthermore, their broad deployment is limited by the sparsity and uneven coverage in the chemical space of available reaction datasets, constraining the transferability of current generative models to network-level reaction discovery for systems involving diverse intermediates or out-of-domain chemistry.

% Paragraph 5: Our contribution
While automated CRN exploration frameworks~\cite{YARP2023, unsleber_chemoton_2022, Steiner2024, https://doi.org/10.1002/anie.202011941, https://doi.org/10.1002/jcc.26734, https://doi.org/10.1002/wcms.1538} and generative TS models have advanced largely in parallel, their integration into a unified workflow remains largely unexplored, limiting the practical application of generative models to large-scale reaction network discovery. In this work, we introduce ByteCRN, a unified and modular framework for chemical reaction network construction and characterization that embeds generative modeling within a physics-based workflow. ByteCRN couples chemically informed reaction enumeration, reactant-product pair construction, generative TS proposal, and generative reactant-product validation within a single CRN pipeline. 
This design reframes generative models from a TS-guessing tool into a network-level component that both screens candidate reactions and validates their connectivity. 
We first establish the accuracy of our generative TS-generation and validation modules on a refined Transition1x dataset, showing improved reaction recovery rate and chemical validity of the generated TS structures relative to conventional quantum chemistry approaches and existing generative models. 
We then demonstrate full CRN construction and characterization on two simple benchmark CRN systems: cyanoacetaldehyde, where ByteCRN recovers the established network while identifying additional reaction channels at a fraction of the cost and $\gamma$-ketohydroperoxide (KHP), a more complex CRN out of the training data distribution featuring dense multichannel reactivity.
%validated by Density Functional Theory level computations, and $\gamma$-ketohydroperoxide (KHP), a more complex CRN featuring dense multichannel reactivity and lying outside the training domain. 
These results establish ByteCRN as a scalable route to automated CRN exploration that substantially reduces computational cost while remaining physically grounded and chemically meaningful. 

\clearpage
\section{Results}
\begin{figure}[!htbp]
    \centering
    \includegraphics[width=0.95\linewidth]{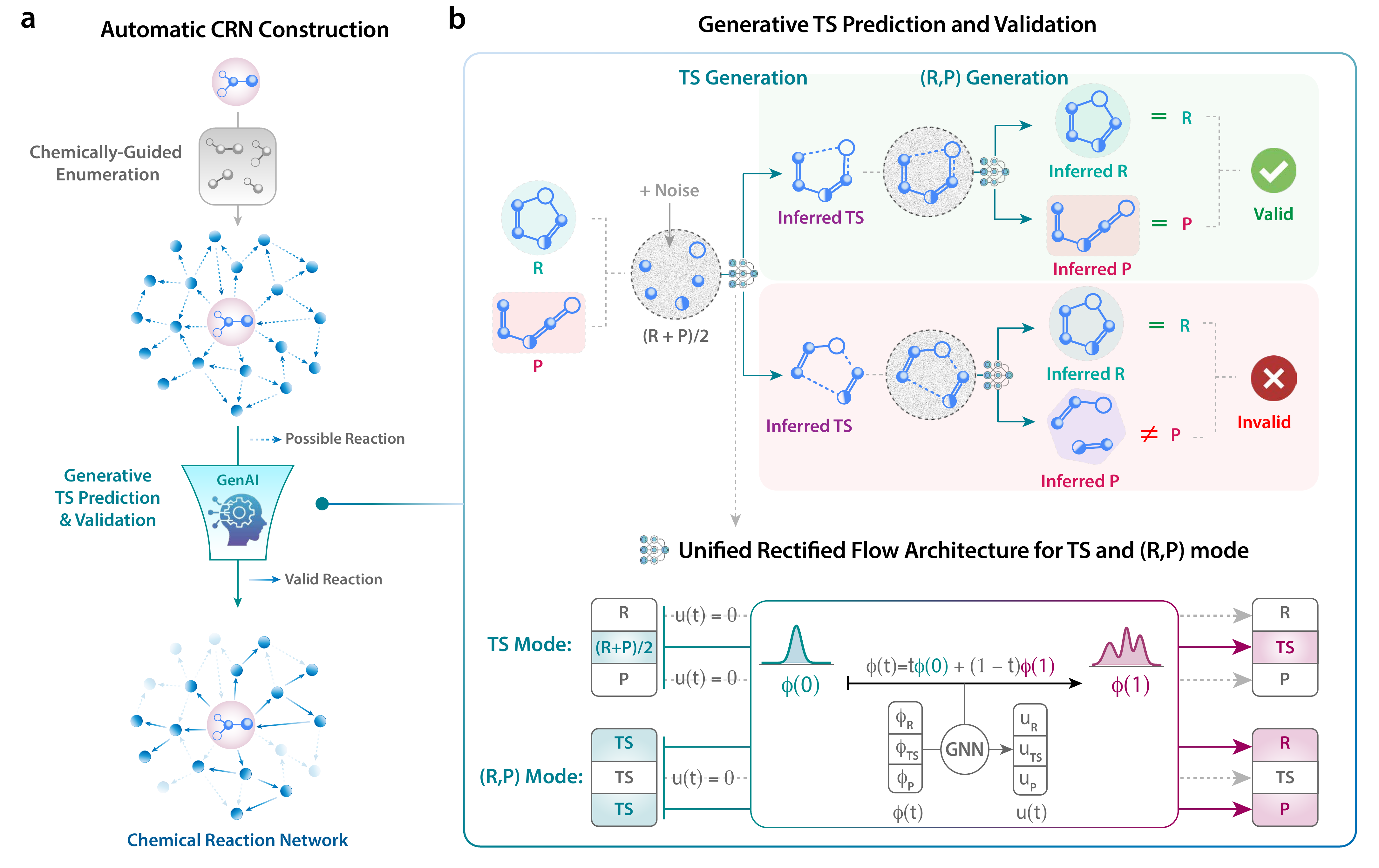}
    \caption{\textbf{ByteCRN workflow for generative chemical reaction network construction}: \textbf {(a)} Starting from the root node (initial molecules), chemically informed enumeration based on inexpensive physical methods (force field or tight-binding) are used to generate candidate reactant--product pairs, which are organized into reaction channels (conformation-dependent elementary reaction routes) and passed to the generative TS prediction and validation module. Validated channels are assembled into the final CRN containing only kinetically valid pathways. \textbf{(b)} The generative TS prediction and validation module. Upper panel: Reactants (R) and products (P) are used to generate an inferred TS by forward inference from their geometric midpoint with some added noise. The inferred TS is then used as the initial state to reconstruct R and P. A reaction is considered valid only if the recovered endpoints match the original pair. Both TS and (R,P) generation modes adopt a rectified-flow architecture acting on 3D geometries of the \((\text{R}, \text{TS}, \text{P})\) triplet, with an E(3)-equivariant graph neural network backbone (LeftNet~\cite{LEFTNET2024} in this study). The TS and RP generators are trained independently and deployed sequentially during CRN construction. }
    \label{fig:schema}
\end{figure}
\FloatBarrier

As illustrated in Figure~\ref{fig:schema}a, the ByteCRN framework is comprised of four main stages, (i) reaction enumeration based on bond breaking and forming, (ii) conformation sampling, (iii) automated reaction prediction via generative models, and (iv) subsequent kinetic calculations. We detail ByteCRN workflow in Appendix~\ref{app:workflow}.
% While automated chemical reaction prediction is a comprehensive challenge, step (iii) is widely recognized as the most computationally intensive stage, significantly hindering progress compared to the other stages (i, ii, iv). We address this long-standing bottleneck by employing generative models.
Our automated reaction prediction and validation approach is motivated by the insight that the TS encapsulates the majority of the reaction pathway information. Theoretically, the TS corresponds to a first-order saddle point on the potential energy surface, representing the highest energy point along the minimum energy path (MEP) that connects the reactants and products. This path is traditionally mapped via IRC validations, which descend from the TS toward local minima by following the eigenvector associated with the unique imaginary frequency.~\cite{fukui_path_1981, gonzalez_reaction_1990} Consequently, the sequential workflow of TS search and IRC calculation can be viewed as an information encoding and decoding process, analogous to a variational autoencoder (VAE).~\cite{VAE2013, kim_diffusion-based_2024} From this perspective, the TS serves as the physical latent space---a low-dimensional bottleneck that compresses the high-dimensional MEP into the most critical degrees of freedom necessary to reconstruct the entire reaction pathway.

Accordingly, we replace these computationally intensive stages with two coupled rectified flow models. As shown in Figure~\ref{fig:schema}b, the first model encodes reaction information from the reactants (R) and products (P) to the TS, while the second decodes from the inferred TS back to its corresponding R and P. A valid TS is characterized by mathematical properties that ensure the derived R and P distribution is narrowly focused around the initial R and P distribution. This approach allows us to jointly determine the TS structure and the reactivity between candidate reactants and products, thereby automating reaction prediction and accelerating CRN generation. We assess reconstruction quality by comparing SMILES strings between original and reconstructed R-P pairs to verify their matching performance. 

We use rectified flow~\cite{REACTIFIEDFLOW2022} to model the continuous molecular transformations among R, TS, and P. In all cases, the neural network takes the joint reaction representation (R,TS,P) as input, but different subsets of the triplet are allowed to evolve depending on the generation mode. In TS mode, the R and P coordinates are fixed by setting their velocities to zero, so that only the TS coordinates evolve from the initial state toward a candidate structure. In RP mode, the TS coordinates are fixed by setting the TS velocity to zero, while the R and P coordinates evolve to reconstruct the endpoint structures. This mode-specific velocity masking enables bidirectional generation within a single rectified-flow architecture. The simplicity of rectified flow makes the model more robust and allows it to learn chemical intuition more intrinsically.  E(3) equivariance and object-aware symmetry are guaranteed by using LeftNet,~\cite{LEFTNET2024} an equivariant graph neural network, and blocked message passing,~\cite{OAREACTDIFF2023} respectively. Random perturbations are added to the initial states during training to further reduce the inductive bias and improve robustness. This dual-mode design ensures that the predicted TS are not only structurally plausible but also chemically consistent with the corresponding R and P, forming the core of ByteCRN's automated reaction prediction pipeline.

\subsection{Transition State and Reaction Prediction}\label{subsec:reaction_prediction}

\begin{figure}[!t]
    \centering
    \includegraphics[width=0.9\linewidth]{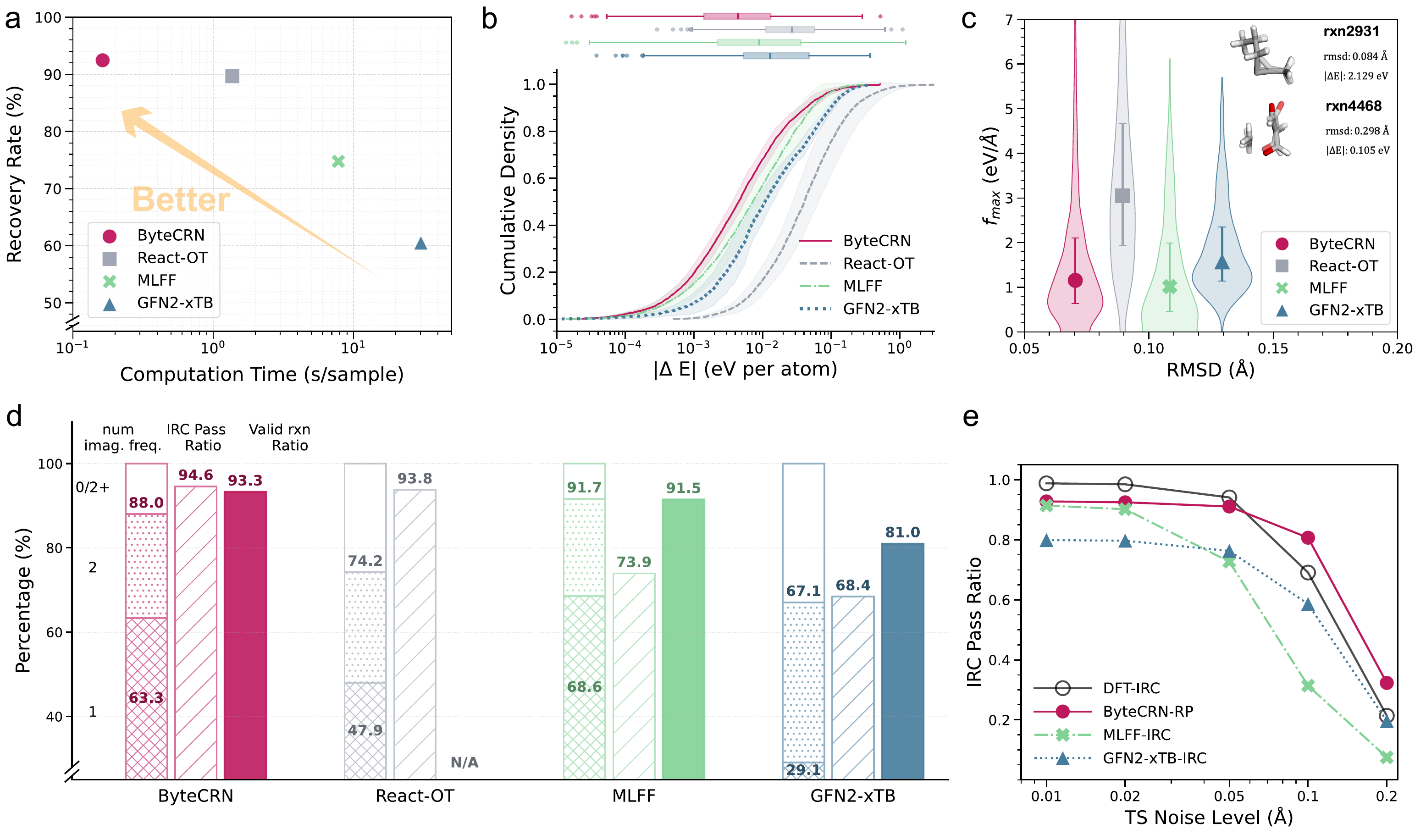}
    \caption{
        \textbf{Performance of ByteCRN on reaction prediction.}
        \textbf{(a)} Reaction recovery rate and runtime comparison for ByteCRN, React-OT, MLFF, and GFN2-xTB through TS search and IRC-like validation.
        \textbf{(b,c)} Absolute TS energy errors and maximum atomic forces (\(f_{\text{max}}\)) relative to DFT-refined TS structures.
        % demonstrates the limitations of RMSD as a standalone metric for comparing TS structures, since TS structures can have a low RMSD but a large maximum force, arising from incorrect bond formation in the predicted TS; and vice versa, where structural discrepancies are confined to non-reactive atoms irrelevant to the key reaction coordinate.
        \textbf{(d)} TS discovery and validation accuracy of ByteCRN-TS and ByteCRN-RP against generative and traditional baselines.
        \textbf{(e)} Robustness of IRC-like validation under varying TS structural noise (0.01--1.0 $\text{\AA}$).
        % : IRC pass ratio is plotted against noise amplitude, with ideal methods matching the trend of DFT-IRC benchmarks. 
        }
    \label{fig:main_figure_ai_res}
\end{figure}

To rigorously evaluate ByteCRN under chemically verified reaction context, we first established a high-quality benchmark by refining the widely used Transition1x dataset.~\cite{TRANSITION1X2022} Our preliminary analysis revealed that approximately 10\% of the entries in the original dataset failed IRC validation, implying that the labeled TS structures did not physically connect the corresponding reactants and products. Training or evaluating on such data would fundamentally misrepresent a model's ability to map valid reaction pathways. Consequently, we curated a ``refined Transition1x'' subset where every reaction is strictly validated via IRC calculations. Our refinement workflow included TS saddle point optimization, followed by IRC validations with the quasi-Newton method in Sella~\cite{SELLA2022}; all calculations were run at the $\omega$B97x/6-31G* level of theory to match the original Transition1x computational setup, yielding a final set of 8,843 reactions from the full dataset's 10,076 raw entries. All density functional theory (DFT) calculations were carried out using gpu4pyscf.~\cite{GPU4PYSCF2024a,GPU4PYSCF2024b}  

On this clean benchmark, we compared ByteCRN against state-of-the-art baselines, including the generative model React-OT~\cite{REACTOT2024}, machine learning force fields (MLFF)~\cite{ZHAOQIYUAN2025}, and semi-empirical GFN2-xTB method~\cite{GFN2XTB2019}-based TS search and IRC validations. We adopted the data split protocol from Ref.~\citenum{REACTOT2024}, but excluded or corrected the aforementioned IRC-failed reactions. To ensure a fair comparison, React-OT was retrained from scratch on our refined dataset. For the MLFF baseline, we employed LeftNet as the architectural backbone, training it on Transition1x energies and forces extracted from the original TS search records; this training was restricted to our refined training partition to prevent any data contamination. TS searches for both MLFF and GFN2-xTB were conducted via the Nudged Elastic Band (NEB) method, following the procedure established in NeuralNEB \cite{NEURALNEB2022}. Finally, the IRC validation for these baselines utilized the same pipeline as our data refinement, substituting the reference DFT calculations with respective MLFF or GFN2-xTB energy and force evaluations.

%The primary barrier to scaling CRN construction is the trade-off between computational cost and reaction recovery rate. In a dense reaction network, missing a single low-barrier pathway can fundamentally alter the predicted product distribution. As shown in Figure~\ref{fig:main_figure_ai_res}a, ByteCRN achieves a superior reaction recovery rate of 92.46\% with an average inference time of just 0.12 seconds. This efficiency is transformative: while traditional density functional theory (DFT) calculations (taking hours per reaction for reactions in Transition1x using gpu4pyscf), and its surrogate MLFF/GFN2-xTB achieves seconds to minutes with scarifce of precision, ByteCRN enables the comprehensive screening of thousands of channels in real-time. This capability ensures that the generated CRN is not just a sparse skeleton of major pathways but a dense, kinetically complete map of the chemical space.

The primary barrier to scaling CRN construction is the inherent trade-off between computational cost and precision. We define the reaction recovery rate as the percentage of test-set reactions successfully processed through the full prediction pipeline. For the React-OT baseline -- which lacks an intrinsic IRC calculator -- we includes MLFF-based IRC validation.  In a dense network, missing even a single low-barrier pathway can fundamentally alter the predicted product distribution. As illustrated in Figure~\ref{fig:main_figure_ai_res}a, ByteCRN achieves a superior reaction recovery rate of 92.46\% with an average inference time of just 0.12 seconds. While DFT calculations---even when accelerated by gpu4pyscf---require hours per reaction, and surrogate MLFF/GFN2-xTB methods require seconds to minutes at the expense of precision decrease at most ~40\%, ByteCRN enables the comprehensive screening of thousands of channels in real-time. This capability ensures that the resulting CRN is not merely a sparse skeleton of major pathways, but a dense, kinetically complete map of the chemical space.

In CRN construction, the accuracy of TS structures is critical, as it directly influences barrier heights and network kinetics. Traditionally, Root-Mean-Square Deviation (RMSD) has been the primary metric for evaluating TS structures. Shown in Figure~\ref{fig:main_figure_ai_res}c, generative models show lower RMSD compared to MLFF/GFN2-xTB methods as those surrogate models may not capture the full geometries in potential energy surface (PES) around reaction center. However, we note that molecular properties fluctuate sharply during bond-forming and breaking phases, where even minor spatial deviations in the reaction center can lead to significant energy discrepancies (Figure~\ref{fig:main_figure_ai_res}c). To address this, we evaluate TS quality using more physically rigorous metrics including: (1) the absolute energy errors of TS per atom, $|\Delta \mathrm{E}|$ ;  (2) the maximum atomic forces $f_{\text{max}}$ ; (3) the number of imaginary frequencies in the Hessian; (4) the success rate of IRC validations at the DFT level.

We observed that while React-OT exhibited a significantly lower RMSD compared to MLFF/GFN2-xTB, it performed poorly in terms of $|\Delta \mathrm{E}|$ and $f_{\text{max}}$. This discrepancy stems from the linear interpolation from a reactant-product midpoint for initialization. This approach imposes an excessive inductive bias on the model, causing it to saturate quickly on training data while failing to capture more general chemical intuition. ByteCRN-TS addresses this limitation by introducing random Gaussian noise to the midpoint initialization. As a result, it achieves superior performance across RMSD, $|\Delta \mathrm{E}|$, and $f_{\text{max}}$ compared to baseline methods, as illustrated in Figures \ref{fig:main_figure_ai_res}b-d and Appendix~\ref{app:training}. Notably, generative models outperform MLFF/GFN2-xTB in vibrational analysis and IRC validations, with ByteCRN-TS achieving an 88\% success rate in frequency analysis ($<2$ imaginary frequencies) and 94.6\% in IRC validation. Conversely, MLFF fails in IRC validation despite reasonable vibrational results, suggesting it struggles to distinguish subtle PES variations across different reactions.

During CRN exploration, reaction enumeration inevitably generates a vast number of invalid reaction candidates. Although generative models excel at TS structure prediction, they cannot independently filter out these invalid entries. This limitation has hindered the integration of previous works~\cite{TSGAN2021,OAREACTDIFF2023,REACTOT2024,TSDIFF2024} into end-to-end CRN discovery pipelines. Traditional IRC methods are notoriously computationally demanding---requiring expensive Hessian evaluations---and remain highly sensitive to hyperparameters. To address this, we leverage the property of the TS as an information bottleneck and introduce ByteCRN-RP, a generative model mapping from TS to RP, as a high-efficiency surrogate.

Our results demonstrate that ByteCRN-RP achieves a 93.3\% success rate, outperforming both MLFF (91.5\%) and GFN2-xTB (81.0\%). To evaluate the ability of the IRC to distinguish invalid reactions, we systematically perturbed the original TS structures in the Transition1x dataset by introducing varying levels of Gaussian noise. Our experiments on these noisy datasets reveal that MLFFs are significantly less robust, likely failing to capture under-sampled regions of the configurational space. In contrast, ByteCRN-RP closely mirrors DFT-level IRC trends, demonstrating a superior capability to accurately distinguish between valid and invalid TS structures.

\FloatBarrier
\subsection{Benchmarking ByteCRN Performance on Network-Level.}

Having established that ByteCRN-TS and -RP achieve state-of-the-art reaction recovery with minimal accuracy loss relative to DFT calculations, we now evaluate the full framework's capacity for constructing and characterizing complete CRNs. We benchmark performance using two molecular systems with distinct roles. First, cyanoacetaldehyde from the Transition1x dataset serves as an in-domain test to evaluate precision-accuracy trade-offs against established GFN2-xTB-based workflows.~\cite{YARP2021,YARP2023} Second, we examine $\gamma$-ketohydroperoxides (KHPs), which present a more rigorous challenge due to their densely branched, multi-channel reaction networks. Crucially, embedding analysis reveals that the KHP system lies outside our training distribution, providing an ideal test bed for assessing the model's ability to generalize in out-of-domain scenarios.

\begin{figure}[!t]
\centering
\includegraphics[height=0.55\textheight,width=\linewidth,keepaspectratio]{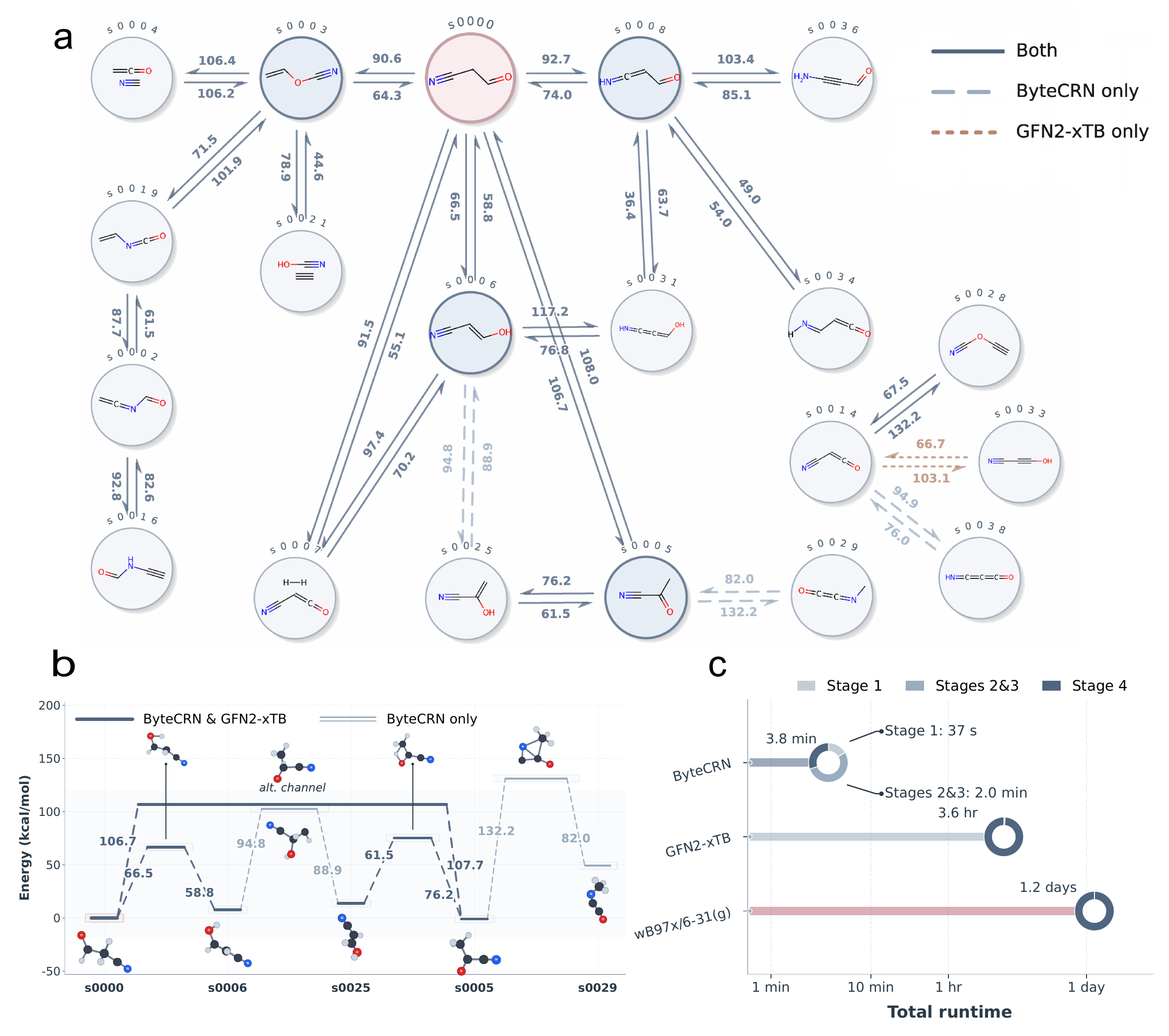}
\caption{\textbf{Reaction network and representative pathway of cyanoacetaldehyde.}
\textbf{(a)} DFT-refined CRN generated by ByteCRN and the GFN2-xTB workflow; dark blue, light grey, and orange edges denote shared, ByteCRN-only, and GFN2-xTB-only reactions, respectively; edge labels give activation barriers in kcal mol$^{-1}$.
\textbf{(b)} Energy profile and optimized TS structures along the highlighted s0000 $\rightarrow$ s0006 $\rightarrow$ s0025 $\rightarrow$ s0005 $\rightarrow$ s0029 pathway.
\textbf{(c)} Stage-wise runtime comparison for ByteCRN and conventional GFN2-xTB workflows. Stage 1 denotes reaction enumeration, Stages 2--3 denote conformer generation and reaction-pair preparation, and Stage 4 denotes TS proposal, refinement, and connectivity validation.}
\label{fig:3cyanopropanal_crn}
\end{figure}

For cyanoacetaldehyde, ByteCRN identifies a more diverse array of reaction pathways within minutes, outperforming GFN2-xTB workflows (3.5 hours) and DFT-level methods (1.2 days), as shown in Figure~\ref{fig:3cyanopropanal_crn}a and c. To simulate research environment of individual researchers on personal PCs, benchmarks were performed using an NVIDIA 4090 for generative models and 32-core CPUs for GFN2-xTB. While reaction enumeration and conformer sampling are computationally inexpensive, the reaction prediction stage remains the primary bottleneck for traditional methods, typically requiring hours or days. ByteCRN leverages generative models to accelerate exploration at this stage, effectively addressing this long-standing computational hurdle.

As discussed in Section~\ref{subsec:reaction_prediction}, ByteCRN's generative models effectively capture the underlying chemical intuition of reactions. In contrast, GFN2-xTB may fail to map complex variations of the PES, leading to diminished predictive performance. This advantage allows ByteCRN to discover additional reaction channels with more intermediates and lower energy barriers compared to GFN2-xTB-based workflows (Figure~\ref{fig:3cyanopropanal_crn}b). 
We note that this test case should not be taken to imply that ByteCRN always discovers novel reactions or lower-barrier pathways in all cases, as its performance depends significantly on the training data. However, as data coverage expands across diverse chemical space, the domain of applicability of ByteCRN will broaden continuously, enabling robust performance in the discovery of complex reaction networks. We next demonstrate its potential in the more complex KHPs system.

\FloatBarrier
\pagebreak[3]

 3-hydroperoxypropanal is the simplest member of the KHP family (hereafter referred to as KHP). As a prototypical $\gamma$-KHP, it combines peroxide and carbonyl functionalities within a single molecular framework and undergoes rich unimolecular chemistry, including the canonical Korcek mechanism. Its structure and reactivity have been extensively characterized at the M05-2X-D3/def2-SVP level of theory~\cite{YARP2023}, providing a well-established reference for evaluating the fidelity of our AI-assisted CRN construction and TS discovery workflow. Given its chemical complexity and limited representation in the training data, KHP serves as an ideal testbed for evaluating the out-of-distribution generalization capability of generative reaction models.

\begin{figure}[!t]
\centering
\includegraphics[height=0.58\textheight,width=\linewidth,keepaspectratio]{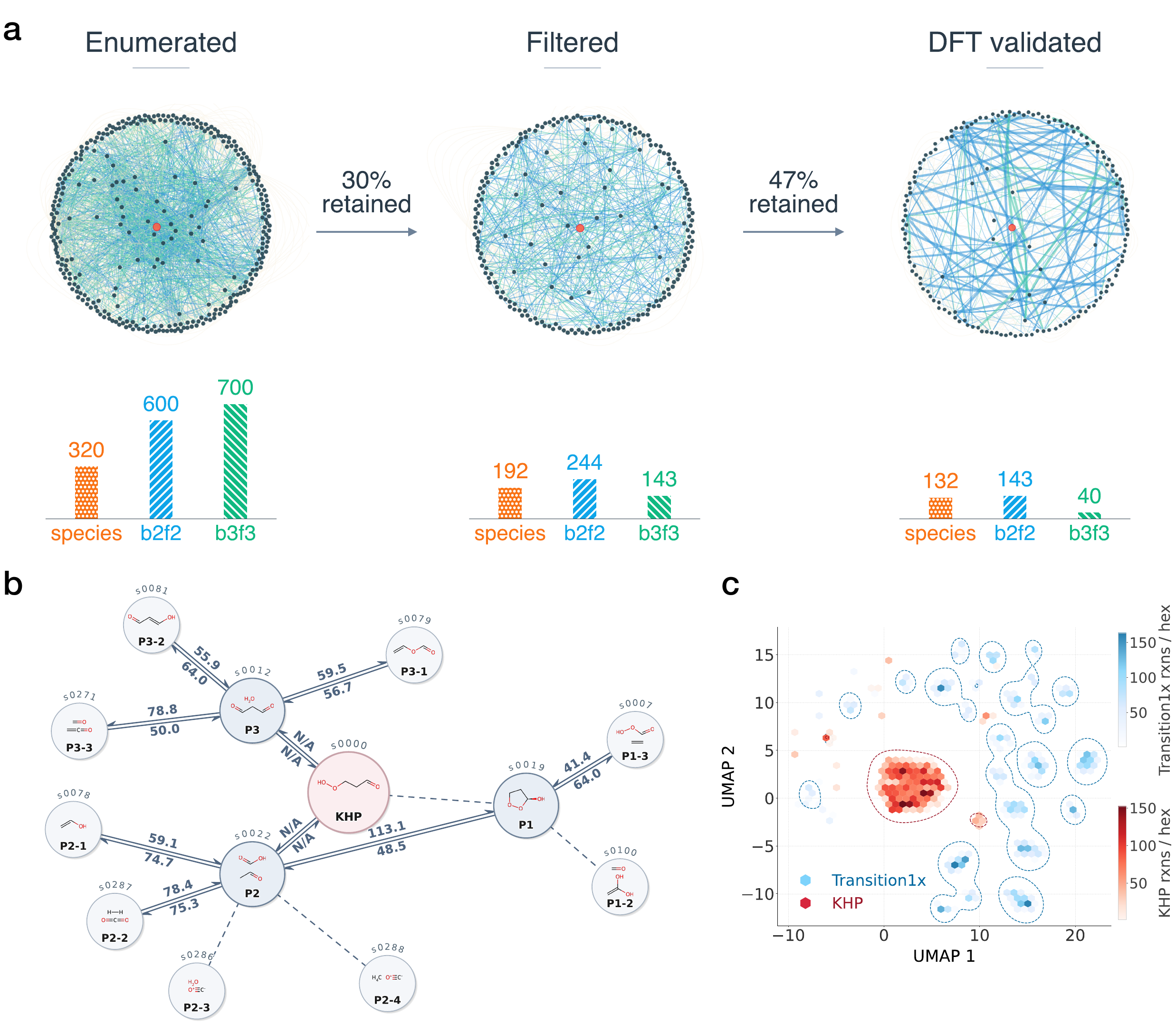}
 \caption{\textbf{Out-of-distribution CRN construction for 3-hydroperoxypropanal (KHP).}
\textbf{(a)} Stage-wise KHP network construction. Graph-based enumeration generates 1{,}300 unique reactions, the ByteCRN generative filter retains 387 candidate reactions, and DFT-level TS optimization, frequency analysis, and IRC validation yield 183 validated reactions. Blue and green edges represent b$_2$f$_2$ and b$_3$f$_3$ reaction channels, respectively. For DFT-validated pathways, edge widths are scaled according to the activation barrier, with thicker edges indicating lower-barrier reactions.
\textbf{(b)} DFT-refined KHP subnetwork compared with the 13 kinetically relevant pathways reported by Zhao et al. 2023~\cite{YARP2023}. Solid edges denote pathways recovered with validated TS structures, while dashed edges mark literature channels not validated by ByteCRN in this run.
\textbf{(c)} UniMol-2 reaction-pair embedding of Transition1x and KHP reactions. The separated KHP cluster indicates that the KHP benchmark lies outside the Transition1x training distribution.
}
\label{fig:KHP_crn}
\end{figure}

Figure~\ref{fig:KHP_crn}a shows the network construction pipeline for KHP. The enumeration step defines the chemical space by generating bond-forming and bond-breaking combinations using b$_2$f$_2$ and b$_3$f$_3$ patterns, yielding about 1,300 unique reactions (6,108 total channels), with more b$_3$f$_3$ (700 reactions) than b$_2$f$_2$ (600 reactions). A generative AI filter retains 387 reactions (30\%). Each reaction is proposed with multiple candidate reaction channels, giving 727 channels passed to DFT. After DFT optimization, frequency checks, and IRC validation, 331 valid TS channels remain, corresponding to 183 unique reactions (47\% retention from the ByteCRN set, $\sim$46\% at the channel level). The reaction types also shift across stages. While enumeration is dominated by b$_3$f$_3$, the final validated set is strongly b$_2$f$_2$-dominant (143 vs 40), indicating that simpler two-bond rearrangements are more often viable for KHP. 

The chemical relevance of the constructed network is further supported by comparison with literature pathways. As shown in Figure~\ref{fig:KHP_crn}b, ByteCRN recovers 9 of the 13 reaction channels reported previously for KHP~\cite{YARP2023}, despite limited exposure to similar reactions during training. These recovered pathways include oxygen--oxygen bond cleavage and hydrogen-transfer steps characteristic of KHP degradation, suggesting that ByteCRN does not merely interpolate within its training distribution but captures transferable chemical reactivity patterns. Combined with DFT-level validation, this result demonstrates that ByteCRN can autonomously construct complex OOD reaction networks with high fidelity.

To quantify the distributional challenge posed by KHP, we further analyzed reaction-pair embeddings using UniMol-2 (1.1B parameters).~\cite{UNIMOL2024} For each reaction, the embedding was defined as the average of the reactant and product embeddings to preserve reactant--product exchange symmetry. As illustrated in Figure~\ref{fig:KHP_crn}c, the KHP-enumerated reactions form an isolated cluster in the 2D UMAP projection~\cite{UMAP2018}, with little overlap with the Transition1x training reactions. The correlation analysis in Appendix~\ref{app:ood_khp} further shows that Transition1x reactions exhibit much stronger self-correlation than cross-correlation with KHP reactions. These results confirm that KHP is not simply another in-distribution test case, but a stringent benchmark for evaluating model generalization. Together, the validated CRN construction, literature-pathway recovery, and embedding analysis demonstrate that ByteCRN can generalize to chemically distinct reaction spaces while retaining physical reliability.

\FloatBarrier
\section{Discussions and Outlook}

% Paragraph 1: Discussion - The information bottleneck in *computational* CRN construction is TS. Thus in our work, we mimic an information encoding-decoding process. [just minor wording changes] 

% Paragraph 2: Discussion - Due to the encoding/decoding, we naturally use generative models, which are more generalizable than MLFFs. Reason... [merge the current paragraphs 3&4]

% Paragraph 3: Discussion - Highlight we shift from single reaction to network level. [current paragraph 2] 

% Paragraph 4: "In summary" paragraph. More details, and clear correspondence with each section in the paper. 

% Paragraph 5: Limitations. Gas-phase only; enumeration bound by b2f2; missing reactions due to OOD; 

% Paragraph 6: Outlook. 1. expand dataset/active learning strategy. 2. surface. 3. extending the encode/decode principle to enhanced sampling/CV problems and solve broader rare-event problems. 

Constructing a CRN is a complex process involving reaction enumeration, structural sampling, and kinetic validation. Because the exhaustive exploration of a CRN is computationally prohibitive, striking a balance between efficiency and accuracy is the central challenge in developing exploration algorithms. To address this, we introduced ByteCRN, an integrated workflow powered by generative models. By reframing traditional numerical optimization as an information encoding and decoding process, the ByteCRN-TS and ByteCRN-RP modules replace computationally heavy steps with generative modeling. This innovation enables the exploration of vast chemical reaction spaces significantly faster than traditional methods, without sacrificing accuracy.

%Unlike previous approaches that focus exclusively on transition state prediction for isolated elementary reactions, ByteCRN integrates generative models into the broader context of end-to-end CRN generation. This advancement exposes the models to more realistic, OOD scenarios. Consequently, while benchmarking against curated datasets remains standard practice, the focus shifts toward the more important task of discovering novel reaction pathways, highlighting model generalization as a critical, real-world challenge.

A key distinction of ByteCRN is that generative models are embedded in an end-to-end CRN generation framework, rather than being applied only to transition state prediction for isolated elementary reactions. This setting more closely reflects the bottleneck in autonomous CRN explorations, where candidate reactions are often novel, sparsely represented in training data, and far from curated benchmark distributions. As a result, the relevant measure of model performance is not only accuracy on standard datasets, but also the ability to generalize to out-of-distribution (OOD) reactions and to reveal previously unknown mechanistic pathways.

The necessity of this generalization is evident when evaluating current alternatives. MLFFs and semi-empirical methods have advanced rapidly and are powerful tools for molecular simulation, yet their reliability can degrade when applied to high-dimensional, rugged PESs outside the domain of their training data. For instance, even when an MLFF is extensively trained on the 9.6 million molecular configurations of the Transition1x dataset, minor OOD deviations yield errors in final predictions. Generative models offer a complementary route. Rather than strictly fitting a rugged and highly sensitive PES, generative models learn a broader ``chemical intuition''. This advantage is rooted in the physical reality that while a calculated PES can fluctuate drastically across different levels of quantum mechanical theory, the underlying geometric structures and core reaction mechanisms are frequently conserved. By prioritizing these structural patterns over volatile energy landscapes, generative models can extrapolate far more robustly than traditional PES-based models.

%In summary, ByteCRN provides a practical, scalable approach to autonomous chemical exploration, validated through the discovery of novel pathways in cyanoacetaldehyde and the successful modeling of the challenging $\gamma$-ketohydroperoxide network. 
The practical value of this strategy is demonstrated by the discovery of novel pathways in cyanoacetaldehyde and by the successful modeling of the challenging $\gamma$-ketohydroperoxide network. These examples show that ByteCRN can serve not only as an accelerator for known reaction classes, but also as a scalable platform for autonomous mechanistic exploration. Looking ahead, while the current study focuses on gas-phase small molecules, ByteCRN holds potential for broader applications by incorporating solvent effects and surface reactions. However, this expansion is currently constrained by the high cost of dataset generation, as well-curated reaction datasets remain relatively small and highly heterogeneous. Furthermore, compared to the nearly infinite scale of chemical space, the domain covered by machine learning datasets will remain limited in the foreseeable future. Consequently, designing smarter data sampling strategies---alongside advancing model architectures to capture general chemical intuition---will be important for long-term development. Ultimately, tackling these challenges paves the way for deeper collaboration between computer scientists and chemists to advance the field.

\clearpage
\section{Method}

\subsection{Reaction Network Construction}
\textbf{Graph-Based Enumeration.} ByteCRN performs reaction enumeration utilizing the graph-based framework of YARP v2.0~\cite{YARP2023}. Molecules are represented as labeled graphs where vertices denote atoms and edges denote bonds. Elementary transformations are encoded via systematic bond-breaking and bond-forming rules (b$n$f$m$, breaking $n$ bonds and forming $m$ bonds), encompassing b1f1 (single-bond rearrangements), b2f2 (two-bond breaking and formation, typical in ring closures), and b3f3 (higher-order multi-bond reorganizations) operations. We adopt these operators for structural enumeration and augment them with additional physical screening layers to remove invalid, redundant, or highly implausible structures.

\textbf{Structural and Thermodynamic Filtering.} To prevent the generation of chemically implausible products, ByteCRN applies structural filters including the rejection of user-specified undesired rings, strict enforcement of atomic valency, and canonical SMILES-based deduplication. To ensure the enumerated network remains physically meaningful at early stages, we introduce an additional thermodynamic screening step. Product states exceeding a user-defined reaction-enthalpy threshold (e.g., $\Delta\mathrm{H} \leq 100\ \text{kJ mol}^{-1}$ for atmospheric or low-temperature systems) are discarded. Reaction enthalpies are rapidly estimated using group-additivity schemes, semiempirical methods (such as GFN2-xTB), or machine learning potentials (such as MACE), reserving higher-level electronic structure refinements for key species. The surviving species are canonicalized, assigned unique identifiers, and logged in a global registry to guarantee cross-iteration uniqueness. This coupled structural--thermodynamic filtering effectively prunes large combinatorial branches, accelerating convergence toward chemically relevant regions of the reaction space.

\textbf{Modular Workflow for Autonomous CRN Construction.} The construction pipeline is organized into three sequential modules: initialization, TS candidate generation, and physical validation. This structure separates the construction of chemically plausible reaction channels from the identification and verification of reaction mechanisms, enabling a flexible and extensible framework. The first component, CRN initialization, constructs a high-recall set of candidate reaction channels. Graph-based enumeration generates possible reactions, which are subsequently converted into three-dimensional structures through conformer sampling at the GFN-xTB level of theory via CREST.~\cite{CREST2024} These are then transformed into reaction-ready reactant--product pairs, providing geometrically consistent inputs for downstream TS identification. The second component, TS candidate generation, is designed as a method-agnostic module. In this work, we employ ByteCRN-TS and ByteCRN-RP to propose and validate TS candidates directly from reactant--product pairs. The final component, refinement and validation, provides a unified physics-based evaluation of all TS candidates through quantum chemistry calculations. The initial TS structures are subjected to consistent saddle-point optimization and IRC validations to verify connectivity. The validated reaction channels are then assembled into a chemically consistent reaction network for downstream analysis.

%Notably, the reliability of TS generation depends critically on the quality of the underlying reactant--product representations. Ensuring reaction-ready geometries is therefore essential for enabling effective TS identification, as discussed in the following section.

\subsection{Data Refinement on Transition1x}

The Transition1x dataset was originally developed by Grambow et al. \cite{GRAMBOW2020} via automated PES exploration at the $\omega$B97X-D3/def2-TZVP quantum chemistry level, encompassing 12,000 organic reactions. It was later refined in Ref.~\citenum{TRANSITION1X2022} by performing NEB calculations at the $\omega$B97x/6-31G(d) level on 10,073 of these reactions, yielding the atomic forces and energies of 9.6 million intermediate structures. Transition1x is well-suited for this study due to its rich collection of intermediate structures and relatively accurate TS structures, which enables a rigorous comparison between MLFF and generative models.

However, as reported in Zhao et al. 2025~\citenum{ZHAOQIYUAN2025}, approximately 10\% of the reactions in the Transition1x test dataset failed IRC validations. Given our goal of comprehensively investigating ML methods for constructing CRNs, we therefore refined the entire Transition1x dataset through additional IRC validations. Specifically, because the transition states in Transition1x were only converged to $\mathrm{F_{max}}<0.05\,\mathrm{eV}/\text{\AA}$ and may still deviate from the saddle points of the PES, we first re-optimized all TS structures using Sella.~\cite{SELLA2022} We then performed vibrational frequency calculations at the same $\omega$B97x/6-31G(d) level to assess TS quality, retaining only structures with exactly one imaginary frequency corresponding to the reaction coordinate and discarding structures with zero or multiple imaginary frequencies, which indicate minima or higher-order saddle points. For the remaining transition states, we carried out IRC validations using the Hessian-guided predictor-corrector method in Sella. The IRC trajectories were propagated in both forward and reverse directions from the transition state until the energy converged to within 0.01 eV of a local minimum. A reaction was considered valid only when both IRC paths reached plausible reactant and product structures consistent with the expected chemistry of the reaction class; ambiguous cases were manually inspected, and reactions were discarded if either IRC path failed to reach a reasonable minimum or connected to unexpected chemical species.

%\begin{itemize}
%    \item Transition State Optimization: Transition states in Transition1x are only converged to $\mathrm{F_{max}}<0.05\,\mathrm{eV}/\text{\AA}$, which may still deviate from the saddle point of the PES. To mitigate this, we re-optimize the TS structures via Sella.~\cite{SELLA2022}
    
%    \item Vibrational Analysis: After re-optimization, we perform vibrational frequency calculations at the same $\omega$B97x/6-31G(d) level to confirm the quality of each TS. A valid first-order saddle point should exhibit exactly one imaginary frequency corresponding to the reaction coordinate. We discard any structures with zero or multiple imaginary frequencies, as they either represent minima or higher-order saddle points. 
    
%    \item Intrinsic Reaction Coordinate Validations: For transition states passing vibrational analysis, we perform IRC validations using the Hessian-guided predictor-corrector method in Sella.~\cite{SELLA2022} We integrate in both forward and reverse directions from the transition state until the energy converges to within 0.01 eV of a local minimum. A reaction is considered valid only if both IRC paths successfully reach plausible reactant and product structures that match the expected chemistry of the reaction class. We manually verify edge cases where convergence is ambiguous, and discard reactions where either IRC path fails to reach a reasonable minimum or connects to unexpected chemical species.
%\end{itemize}

\subsection{Flow Matching}

We implement a rectified flow framework for the generation of transition state candidates (ByteCRN-TS) and validation of reaction pathways (ByteCRN-RP).  Rectified flow is a generative modeling technique that learns to transport a simple initial state into a target data distribution along deterministic, straight-line trajectories. In a chemical context, this can be visualized as continuously morphing an initial structural guess ($z_0$) into a target geometry ($z_1$) over a normalized ``time" variable $t \in [0,1]$, governed by the linear interpolation $z_t = t z_1 + (1-t) z_0$. The neural network is then trained to predict the constant driving vector field $\frac{d z_t}{d t} = z_1 - z_0$ that directs this structural evolution.  For TS generation, we employ the E(3)-equivariant LeftNet architecture, consistent with the React-OT framework. To rigorously enforce elemental and stoichiometric conservation during the generative process, we freeze the node feature updates (e.g., atomic identities and formal charges) during training and inference, restricting the network to exclusively update 3D positional coordinates. Furthermore, to simplify the generation trajectory and improve computational efficiency, we replace the original Schrödinger Bridge schedule used in React-OT with a deterministic straight-line schedule. The input to the LeftNet model comprises the geometric triplets of the reactant, transition state, and product ($R, TS, P$). In standard implementations, the initial state for the generative flow is often defined as the exact geometric midpoint between $R$ and $P$. However, we observed that this deterministic initialization leads to severe overfitting on the training manifold and early saturation during training. To mitigate this, we introduce stochasticity by injecting a Gaussian noise ($\epsilon \sim \mathcal{N}(0, \sigma^2 \mathbf{I})$, $\sigma=0.2$) into the midpoint initialization. This noise-augmented initialization acts as a geometric regularizer, significantly improving the model's generalization capabilities and downstream performance on the test data.

To evaluate the quality of generated TS and discard chemically implausible reactions from reaction enumeration, we introduce ByteCRN-RP, which maps the inferred TS geometry back to its corresponding R and P. In this reverse-mapping setting, both the R and P geometries are initialized directly from the inferred TS structure. Reaction directionality is strictly predefined by thermodynamic constraints: higher-energy conformations are manually designated as reactants, while lower-energy conformations are designated as products. To achieve better structural regularization and penalize physically unrealistic geometric distortions, we train the ByteCRN-TS and ByteCRN-RP model using an $\mathrm{L}_1$ loss objective rather than standard mean squared error.

\subsection{Machine Learning Force Field}

Our machine learning force field is built from LeftNet~\cite{LEFTNET2024}, a equivalent graph neural network that can efficiently encode the 3D structure of molecules. LeftNet balances its computational efficiency and expressiveness through local substructure encoding and frame transition encoding. We use the same data split and same architecture hyper-parameters in OA-ReactDiff and React-OT to train a universal potential. Per-atom energy is output using the final node features and a predictive head, and these per-atom energies are summed to get the per-structure energy. Forces are computed as the negative gradient of the per-structure energy with respect to atomic positions. We use the Huber loss with $\delta = 0.01$ to supervise both per-structure energy and per-atom forces, with a 1:20 ratio between the two loss terms.~\cite{ZHAOQIYUAN2025} We use an AdamW optimizer with a learning rate $5e^{-4}$.

\FloatBarrier
\newpage
\bibliographystyle{unsrt}
\bibliography{main}

\FloatBarrier
\newpage
\beginappendix

\section{Transition States as Ideal Latent Space of MEP}
\label{app:latent_ts}

In this study, modeling traditional TS searches and IRC validations offers practical advantages, including an optimal balance between inference speed and accuracy, and a streamlined training process utilizing only positive reaction pairs. However, this approach may appear less intuitive than traditional numerical methods. To address this, we interpret the entire process through the lens of information theory, framing it as a sequence of information encoding and decoding.

A perfect transition state is formally defined as a first-order saddle point on the PES, characterized by exactly one imaginary vibrational mode that corresponds to the reaction coordinate. Conversely, an imperfect transition state may manifest in several ways: (1) the presence of near-zero eigenvalues; (2) multiple negative eigenvalues; or (3) the absence of negative eigenvalues within the Hessian matrix. Our motivation stems from the observation that a perfect transition state uniquely determines the entire MEP via gradient descent along the reaction coordinate. In contrast, an imperfect transition state yields multiple potential solutions or none at all, thereby introducing mathematical ambiguity.

We define $\chi \subset \mathbb{R}^{3N}$ as the configuration space and $\gamma(s) \in \mathcal{P}$ as a path, represented by a continuous map from the reactant $\mathbf{R}$ to the product $\mathbf{P}$, such that $\gamma: [0, 1] \to \chi$. The Encoder ($\mathcal{F}$): Maps the entire path $\gamma$ to a single configuration $x^\dagger$ (the transition state): $\mathcal{F}: \mathcal{P} \to \chi, \quad \mathcal{F}(\gamma) = x^\dagger$. The decoder $\mathcal{G}$ aims to recover the full path from the single configuration: $\mathcal{G}: \chi \to \mathcal{P}, \quad \mathcal{G}(x^\dagger) = \hat{\gamma}$.  In traditional pipelines, $\mathcal{F}$ represents TS search methods (e.g., Nudged Elastic Band); in our framework, it is the ByteCRN-TS generative model. In traditional computational pipelines, $\mathcal{F}$ represents TS search methods such as the NEB method; in our framework, it is embodied by the ByteCRN-TS generative model. Similarly, while traditional workflows utilize IRC methods for $\mathcal{G}$, our approach employs the ByteCRN-RP generative model.

In the information theory perspective, $x^\dagger$ can be viewed as a good representation of the MEP if the mutual information is equal to the entropy of the path itself: $I(\gamma;x^\dagger)=H(\gamma)$. Here, $H(\gamma)$ represents the total uncertainty or information content inherent in the reaction trajectory. The condition $I(\gamma;x^\dagger)=H(\gamma)$ implies that the transition state $x^\dagger$ serves as a sufficient statistic for the path $\gamma$, meaning that all geometric and energetic information required to reconstruct the MEP is losslessly compressed into the transition state configuration. In practical chemical systems, this equality suggests that the transition state acts as a fundamental ``bottleneck" through which the reaction dynamics must pass, effectively encoding the causal link between the reactant and product basins.

This theoretical alignment highlights why generative modeling is particularly well suited for this task, especially when dealing with the three types of imperfect transition states discussed above. ByteCRN-TS encodes the information contained in a reactant--product pair into a candidate TS, emphasizing the local geometry and bond rearrangements most relevant to the reaction coordinate. ByteCRN-RP then decodes from the inferred TS back toward reactant and product basins. For an elementary reaction, the TS lies on a well-defined MEP connecting the input endpoints, so the decoded structures contain a clear and self-consistent signal of the original R--P pair. For a non-elementary or incorrectly paired reaction, however, the inferred TS is not a sufficient statistic for the proposed path: the decoder may relax toward alternative minima, ambiguous intermediates, or endpoints that do not match the input pair. In this sense, reconstruction consistency provides a practical proxy for whether the proposed TS encodes a physically valid elementary reaction.

This perspective also explains why ByteCRN can avoid explicitly modeling every point along the MEP during inference. Rather than learning a full trajectory distribution, the model learns to identify and validate the bottleneck configuration that determines the connectivity of the path. The TS-generation module proposes this bottleneck, while the RP-generation module tests whether the bottleneck contains enough information to recover the original endpoints. A successful encode--decode cycle therefore indicates that the candidate TS is not merely geometrically plausible, but also chemically connected to the intended reactant and product. This information-bottleneck interpretation provides the conceptual basis for replacing iterative TS search and IRC validation with a pair of coupled generative models in the ByteCRN workflow.

\section{Modular Workflow for Autonomous Chemical Reaction Network Construction}
\label{app:workflow}

This appendix expands the module-level implementation of the autonomous CRN construction workflow summarized in the main text. The purpose of this workflow is to separate chemically broad exploration from expensive kinetic validation: inexpensive graph and geometry operations first generate a manageable set of candidate reactions, while generative TS/RP models and electronic-structure calculations are applied only after the candidate space has been filtered.

\begin{figure}[!htbp]
    \centering
    \includegraphics[height=0.5\textheight,width=0.95\linewidth,keepaspectratio]{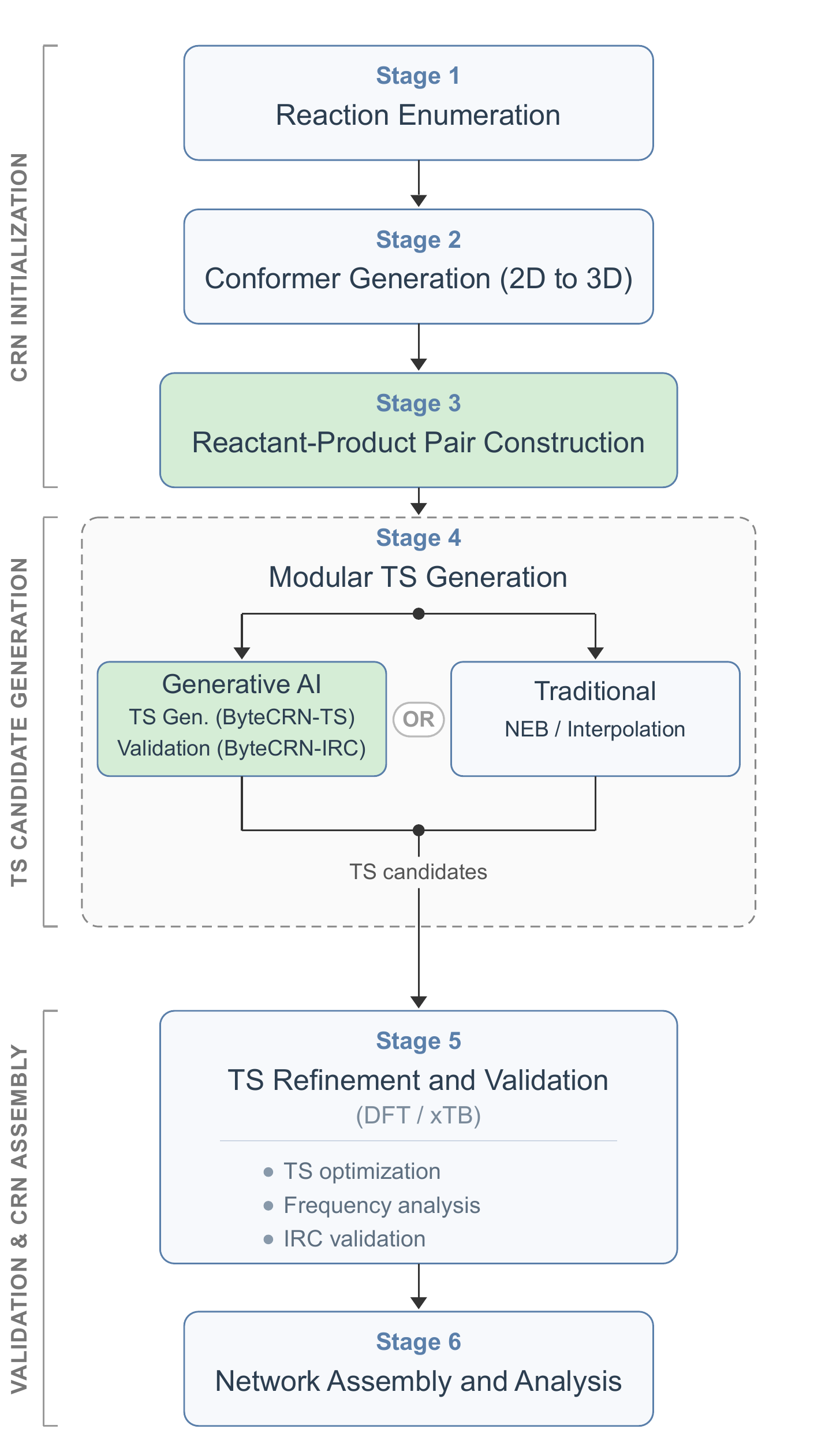}
    \caption{\textbf{Modular workflow for autonomous chemical reaction network construction.} The pipeline consists of reaction enumeration, conformer generation, reaction-ready reactant--product pair construction, modular transition state candidate generation (via either generative AI or traditional methods), and unified transition state refinement and validation, followed by network assembly and analysis.}
\label{fig:bytecrn_workflow}
\end{figure}

As shown in Figure~\ref{fig:bytecrn_workflow}, ByteCRN starts from root species and applies graph-based reaction enumeration to produce candidate products. The resulting species are canonicalized, sampled into three-dimensional conformers, and assembled into reaction-ready reactant--product pairs. Each pair can then be processed either by the generative AI branch, which rapidly proposes TS candidates and validates endpoint connectivity, or by conventional TS-search algorithms when higher-cost reference calculations are required. The final DFT refinement and IRC checks are kept as a unified validation layer so that different upstream proposal methods can be compared within the same network assembly protocol.

This modular design is important for two reasons. First, it makes the workflow extensible: different enumeration rules, conformer generators, TS proposal methods, or validation backends can be exchanged without changing the full CRN assembly logic. Second, it keeps the network chemically auditable, because every retained edge is associated with its originating reactant--product pair, proposed TS channel, refinement status, and final validation outcome.

\section{Data Refinement for Transition1x}
\label{app:data_refinement}

As we use IRC validations to refine the Transition1x dataset, we show the details of the refined data here. The dataset refinement process was critical to ensuring the quality and reliability of our training data, as accurate reaction pathways are essential for developing effective generative models.

The Transition1x dataset underwent rigorous IRC validation to verify the correctness of proposed reaction pathways. This validation process involved calculating intrinsic reaction coordinates from each transition state to confirm that it connects the correct reactants and products. Reactions that failed this validation were either revised or removed from the dataset, ensuring that only chemically feasible reactions were used for training.

\begin{figure}[!htbp]
    \centering
    \includegraphics[height=0.58\textheight,width=0.95\linewidth,keepaspectratio]{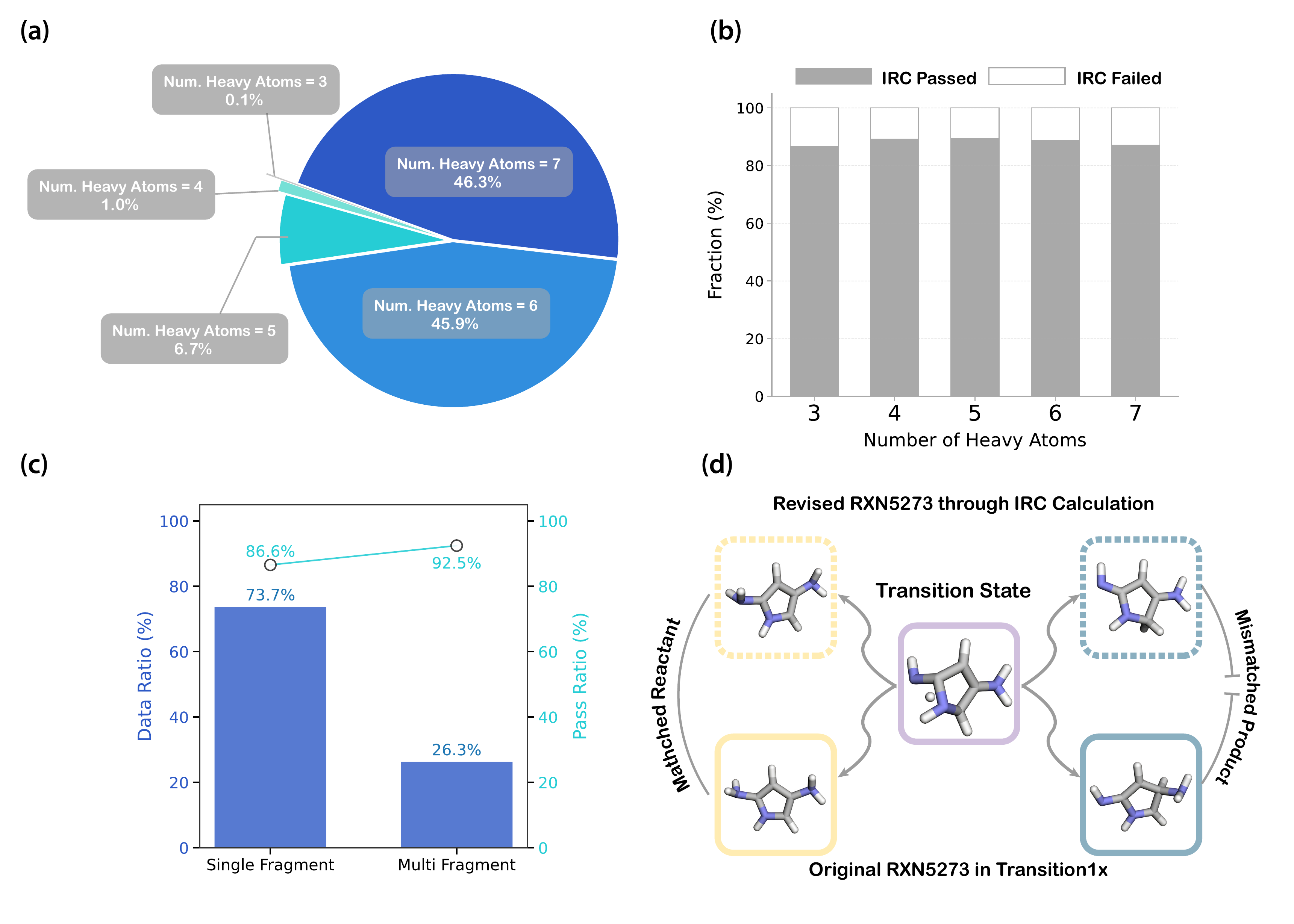}
    \caption{(a) Pie chart showing the distribution of heavy atom counts in the reaction dataset: 7 heavy atoms (46.3\%), 8 heavy atoms (45.9\%), 5 heavy atoms (6.7\%), 4 heavy atoms (1.0\%), and 3 heavy atoms (0.1\%) constitute the dataset; (b) Stacked bar chart quantifying the fraction of reactions that passed (dark gray) or failed (white) IRC validations, binned by heavy atom count (3--7). The majority of reactions passed IRC validation across all heavy atom classes. (c) Bar chart comparing single-fragment and multi-fragment reaction entries: single-fragment reactions account for 85.6\% of the dataset (73.7\% IRC pass rate), while multi-fragment reactions represent 14.4\% (26.3\% IRC pass rate), demonstrating lower IRC success for structurally complex (multi-fragment) systems. (d) Schematic of IRC-guided revision for reaction RXN5273: IRC validations on the original transition state (labeled ``Original RXN5273 in Transition1x'') reveal a mismatched product (vs. the initial proposed product), illustrating how IRC analysis refines reaction outcome predictions during dataset curation.}
    \label{fig:data}
\end{figure}

As shown in Figure~\ref{fig:data}a, the dataset is predominantly composed of reactions with 7 or 8 heavy atoms (92.2\% combined), with smaller contributions from reactions with fewer heavy atoms. This distribution reflects the balance between computational feasibility and chemical complexity in our dataset design. Figure~\ref{fig:data}b illustrates the IRC validation results across different heavy atom counts. The high pass rates across all heavy atom classes (particularly for reactions with 4-7 heavy atoms) indicate the overall quality of the initial dataset. However, the slight decrease in pass rate for reactions with 7 heavy atoms suggests increased computational challenges for larger molecular systems. The distinction between single-fragment and multi-fragment reactions, shown in Figure~\ref{fig:data}c, reveals an important trend: multi-fragment reactions exhibit significantly lower IRC success rates (26.3\%) compared to single-fragment reactions (73.7\%). This observation highlights the increased complexity of reactions involving multiple molecular fragments, which often require more sophisticated treatment in both computational modeling and dataset curation. Figure~\ref{fig:data}d provides a concrete example of how IRC validation refines reaction pathways. For reaction RXN5273, IRC validations on the original transition state revealed a mismatched product compared to the initially proposed structure. This type of discrepancy is precisely what the IRC validation process is designed to catch, ensuring that only reactions with verified pathways are included in the final dataset.

Overall, the data refinement process significantly enhanced the quality of our training data. By systematically validating reaction pathways through IRC validations, we ensured that our generative models were trained on chemically accurate data, which directly contributed to their robust performance across different DFT levels and molecular complexities.

%\section{Examples for ByteCRN-RP Success on Transition1x bad data}
%We here present which the TS validation framework in ByteCRN successfully correct those failed cases in Transition1x dataset. We show the results here.

\section{Robustness of Generative Models to Different Levels of Theory}
\label{app:robustness}

In ByteCRN's conformer sampling algorithms, we employ the GFN2-xTB force field for conformer generation. These structures were subsequently used directly as inputs for ByteCRN-TS and ByteCRN-RP. Here, we validate that our generative models are robust to variations in DFT levels. The training data for the generative model remains consistent with the main text, utilizing the $\omega$B97x/6-31G(d) level of theory. To assess robustness, we relaxed the reactants and products from the test set using three different methods: (1) GFN2-xTB, (2) MMFF94, and (3) UFF force fields. For GFN2-xTB calculations, we utilized the \texttt{xtb} calculator, while for MMFF94 and UFF force fields, we employed calculators from \texttt{openbabel}.

\begin{figure}[!htbp]
    \centering
    \includegraphics[width=0.9\linewidth]{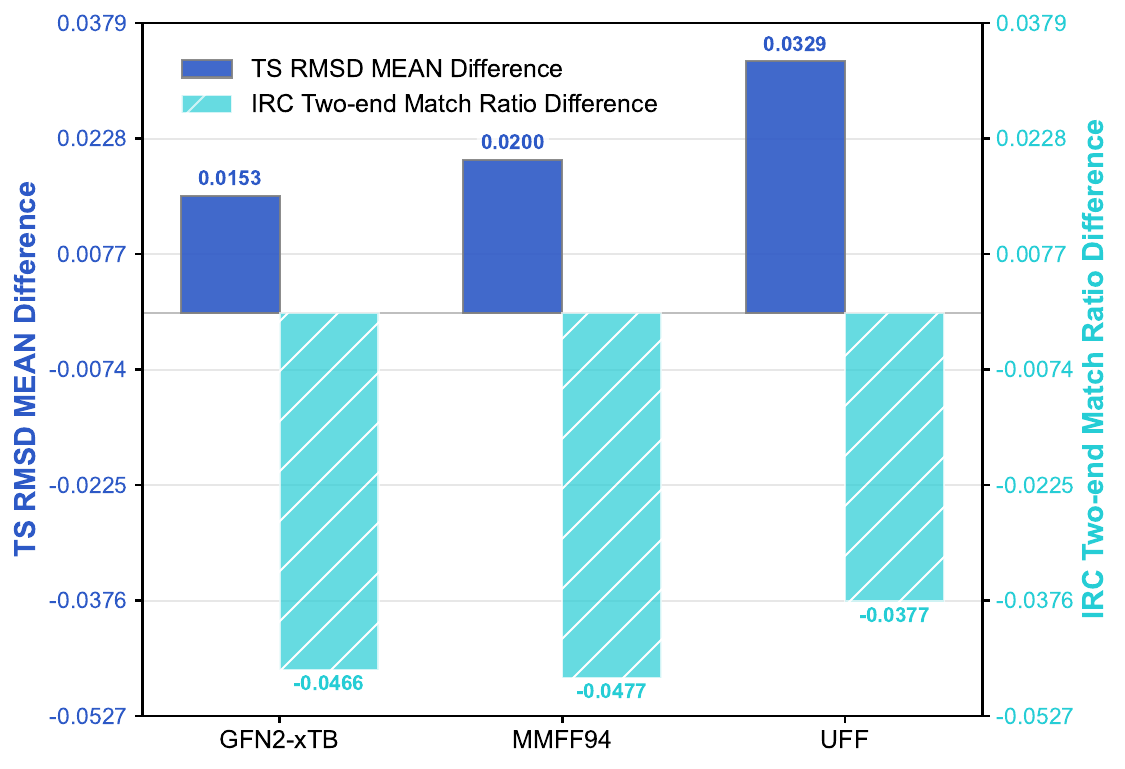}
    \caption{Robustness of ByteCRN to different levels of theory. The figure displays the RMSD values and IRC success ratios when using reactants and products optimized at different levels of theory as inputs to the generative models. Despite variations in optimization methods, the models maintain consistent performance, with RMSD increases and IRC success ratio decreases remaining within acceptable bounds.}
    \label{fig:robustness}
\end{figure}

As shown in Figure~\ref{fig:robustness}, when varying the DFT level theories, the mean RMSD increases by $0.0153\,\text{\AA}$, $0.0200\,\text{\AA}$, and $0.0329\,\text{\AA}$ for GFN2-xTB, MMFF94, and UFF, respectively, compared with inputs optimized at the $\omega$B97x/6-31G(d) level. These values fall within the acceptable range, as demonstrated in Figure~\ref{fig:main_figure_ai_res}h, where RMSD changes of up to $0.05\,\text{\AA}$ do not significantly affect DFT-level IRC validation. Consistent with these observations, the IRC success ratio decreases by at most 4.7\%, further confirming the models' robustness.

Overall, these results demonstrate that our generative models maintain consistent performance across different levels of theory, from advanced DFT methods to classical force fields. This robustness is crucial for practical applications where computational resources may vary, enabling the models to be deployed effectively across different computational environments without significant performance degradation.

\FloatBarrier
\newpage

\section{Training Details}
\label{app:training}

To ensure the reproducibility of experimental results, we present comprehensive training protocols and records in this section. We adopt LeftNet as the inference backbone throughout the entire training procedure, and incorporate an object-aware symmetry constraint into the generative models to enhance structural consistency. All three models are trained from scratch on the refined Transition1x dataset, ensuring a fair and comparable experimental setup. Training are stopped once no performance improvement on test dataset.

\begin{figure}[!htbp]
    \centering
    \begin{subfigure}[t]{0.31\linewidth}
        \centering
        \includegraphics[width=\linewidth]{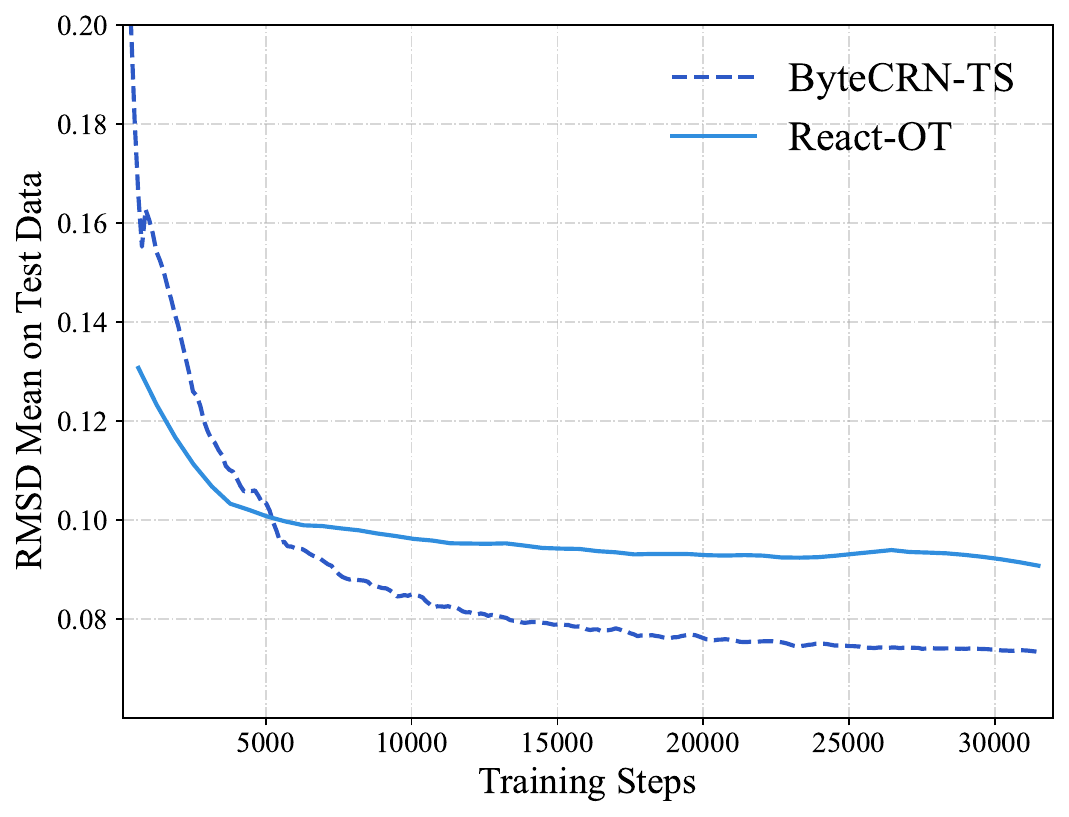}
        \caption{TS RMSD mean}
        \label{subfig:rmsd-mean-genai}
    \end{subfigure}
    \hfill
    \begin{subfigure}[t]{0.31\linewidth}
        \centering
        \includegraphics[width=\linewidth]{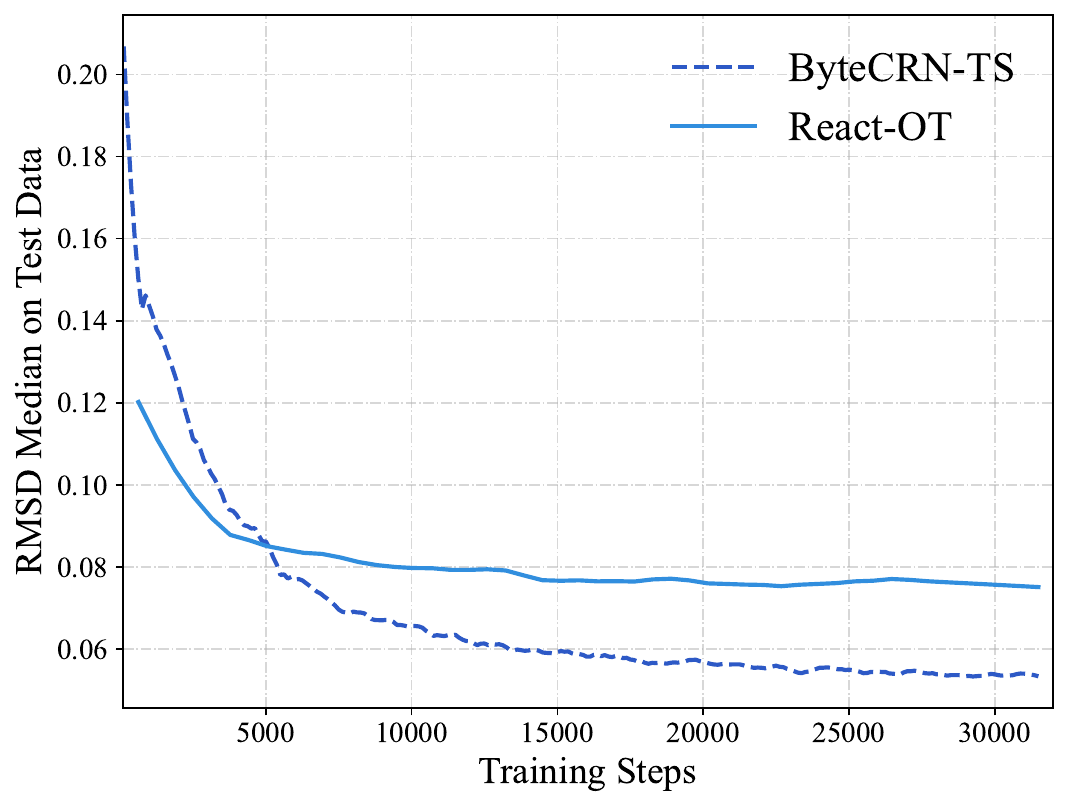}
        \caption{TS RMSD median}
        \label{subfig:rmsd-median-genai}
    \end{subfigure}
    \hfill
    \begin{subfigure}[t]{0.31\linewidth}
        \centering
        \includegraphics[width=\linewidth]{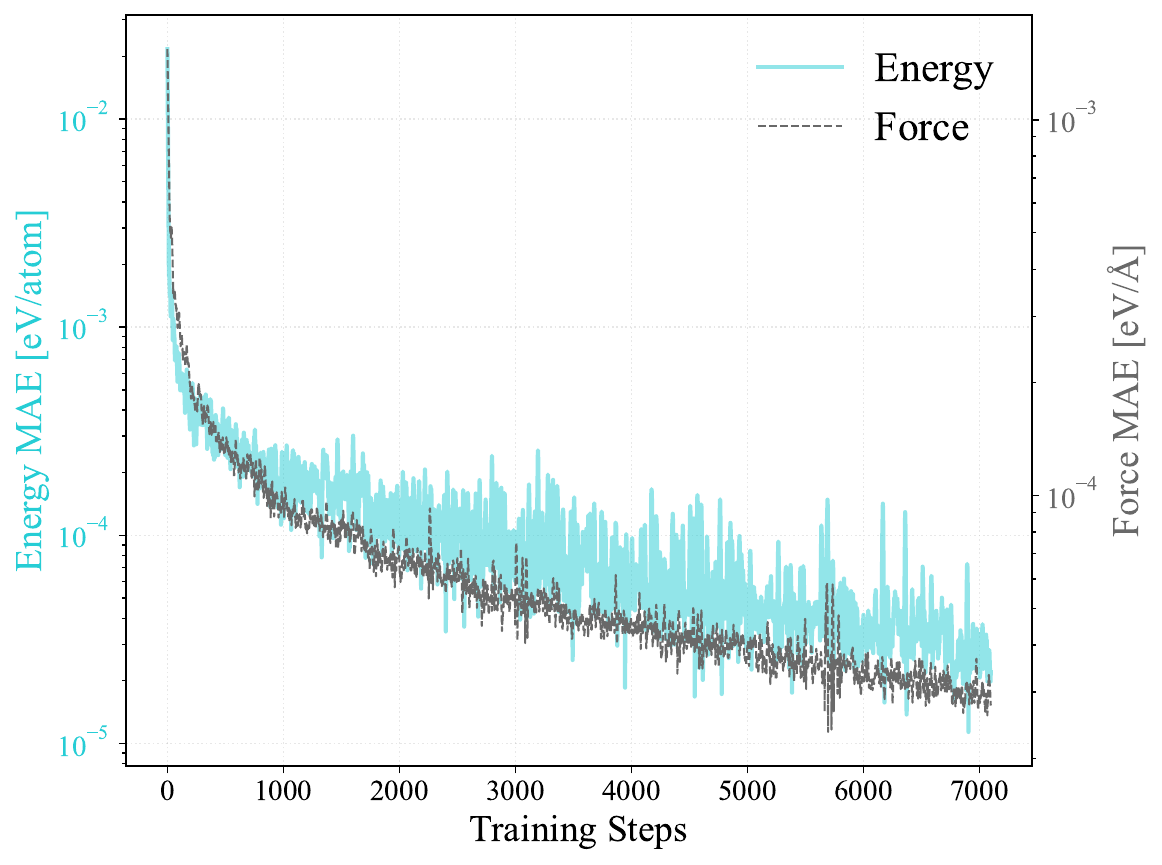}
        \caption{LeftNet energy/force}
        \label{fig:learning_curve_leftnet}
    \end{subfigure}
    \caption{Training curves for the generative transition-state models and the machine learning force field. ByteCRN-TS and React-OT are trained on the refined Transition1x dataset; React-OT converges faster but saturates at a suboptimal endpoint compared to ByteCRN-TS.}
    \label{fig:learning_curve_gen_AI}
\end{figure}

\section{Out-of-Distribution Analysis of the KHP Benchmark}
\label{app:ood_khp}

To quantify the distribution shift between the Transition1x training set and the KHP reaction network, we compared reaction-pair embeddings in the same UniMol-2 representation space used in the main text. We computed nearest-neighbor distances in this embedding space using Transition1x as the reference distribution. The within-Transition1x nearest-neighbor distances define the in-distribution baseline, while the distances from KHP reaction pairs to their nearest Transition1x neighbors measure how far the KHP network lies from the training manifold.

\begin{figure}[!htbp]
    \centering
    \includegraphics[width=0.75\linewidth]{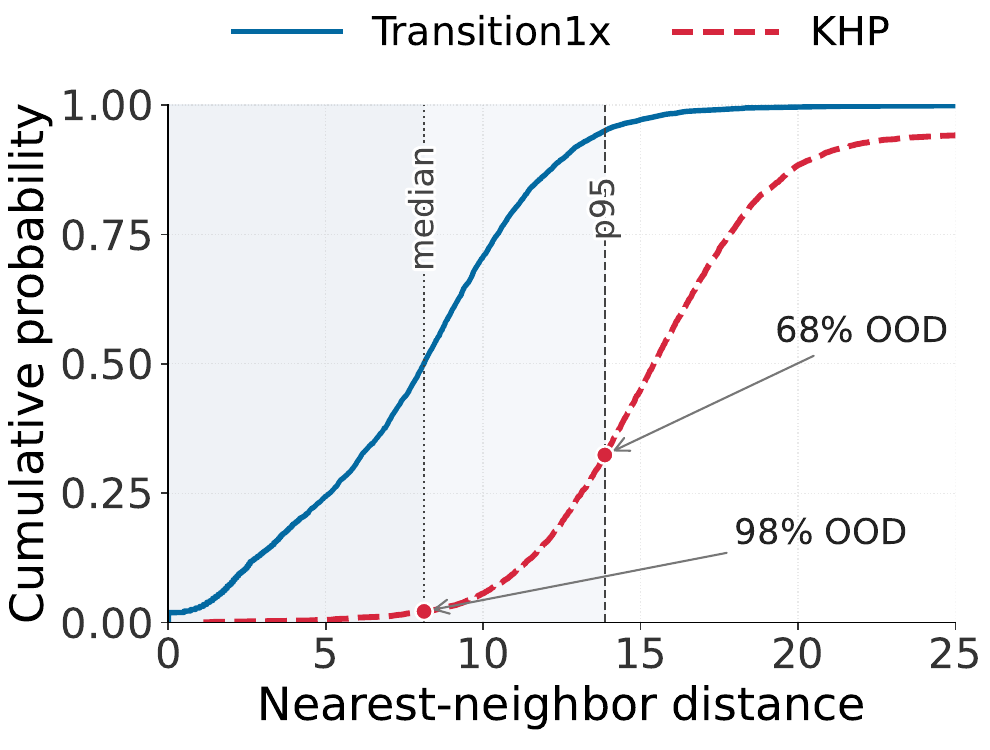}
    \caption{\textbf{Nearest-neighbor embedding-distance analysis for the KHP out-of-distribution benchmark.} The cumulative distributions compare nearest-neighbor distances within Transition1x and from KHP reaction pairs to their closest Transition1x reactions in UniMol-2 embedding space. The Transition1x median and 95th-percentile thresholds define increasingly permissive in-distribution reference regions. The shifted KHP curve shows that most KHP reaction pairs lie far from their nearest Transition1x neighbors, confirming that the KHP CRN benchmark probes out-of-distribution generalization.}
    \label{fig:supp_ood_nn_distance}
\end{figure}

As shown in Figure~\ref{fig:supp_ood_nn_distance}, the KHP nearest-neighbor distance distribution is shifted substantially toward larger distances relative to the Transition1x self-neighbor distribution. Only a small fraction of KHP reactions fall within the Transition1x median nearest-neighbor threshold, and a large fraction remain beyond even the 95th-percentile threshold of the Transition1x baseline. This result supports the main-text UMAP visualization: the KHP benchmark is not simply a denser sampling of familiar Transition1x chemistry, but a chemically distinct reaction space that tests the extrapolative behavior of ByteCRN.

This OOD setting is important for CRN construction because realistic autonomous exploration often proposes reactant--product pairs that are absent from curated training sets. The KHP results therefore evaluate more than interpolation accuracy on benchmark reactions. They test whether the generative TS/RP models can propose chemically useful channels under distribution shift, while the downstream DFT optimization and IRC validation layers retain physical rigor before any reaction edge is added to the final network.

\end{document}